\documentclass{article}
\usepackage{graphicx}  % standard LaTeX graphics tool;
\usepackage{amsmath}   % For getting proper math eqns
\usepackage[compress]{cite}
\usepackage{amssymb}   % Used for getting various symbols eg. gtrsim, lesssim etc.
\usepackage{bm} % For bold math (esp of lower greek letters)
\usepackage{dcolumn}% Align table columns on decimal point
\usepackage{color}
\usepackage{mathrsfs}
\usepackage{amsfonts}
\usepackage{varioref}
\RequirePackage[colorlinks,citecolor=blue,urlcolor=magenta,linkcolor=blue]{hyperref}
\def\EH{Einstein-Hilbert }

\def\gr{general relativity}

\labelformat{section}{Section #1} 
\labelformat{subsection}{Section #1} 
\labelformat{subsubsection}{Section #1}
\labelformat{subsubsubsection}{Section #1}
\labelformat{equation}{Eq.~(#1)} 
\labelformat{figure}{Fig.~#1} 
\labelformat{subfigure}{Fig.~\thefigure#1} 
\labelformat{table}{Tab.~#1} 
\labelformat{appendix}{Appendix #1}
\title{Gravity stabilizes itself}
\author{Sumanta Chakraborty
\footnote{sumantac.physics@gmail.com} and Soumitra SenGupta
\footnote{tpssg@iacs.res.in}\\
{\small{Department of Theoretical Physics, Indian Association for the Cultivation of Science, Kolkata 700032, India}}}
\begin{document}

\maketitle
%%%%%%%%%%%%%%%%%%%%%%%%%%%%%%%%%%%%%%%%%%%%%%%%%%%%%%%%%%%%%%%%%%%%%%%%%%%%%%%%%%%%%%%%%%%%%%%%%%%
%%%%%%%%%%%%%%%%%%%%%%%%%%%%%%%%%%%%%%%%%%%%%%%%%%%%%%%%%%%%%%%%%%%%%%%%%%%%%%%%%%%%%%%%%%%%%%%%%%%
%%%%%%%%%%%%%%%%%%%%%%%%%%%%%%%%%%%%%%%%%%%%%%%%%%%%%%%%%%%%%%%%%%%%%%%%%%%%%%%%%%%%%%%%%%%%%%%%%%%
\begin{abstract}
We show that a possible resolution to the stabilization of an extra spatial dimension (radion) can be obtained \emph{solely in the context of gravitational dynamics} itself without the necessity of introducing any external stabilizing field. In this scenario the stabilized value of the radion field gets determined in terms of the parameters appearing in the higher curvature gravitational action. Furthermore, the mass of the radion field and its coupling to the standard model fields are found to be in the weak scale implying possible signatures in the TeV scale colliders. Some resulting implications are also discussed.
\end{abstract}
%%%%%%%%%%%%%%%%%%%%%%%%%%%%%%%%%%%%%%%%%%%%%%%%%%%%%%%%%%%%%%%%%%%%%%%%%%%%%%%%%%%%%%%%%%%%%%%%%%%
%%%%%%%%%%%%%%%%%%%%%%%%%%%%%%%%%%%%%%%%%%%%%%%%%%%%%%%%%%%%%%%%%%%%%%%%%%%%%%%%%%%%%%%%%%%%%%%%%%%
%%%%%%%%%%%%%%%%%%%%%%%%%%%%%%%%%%%%%%%%%%%%%%%%%%%%%%%%%%%%%%%%%%%%%%%%%%%%%%%%%%%%%%%%%%%%%%%%%%%

%%%%%%%%%%%%%%%%%%%%%%%%%%%%%%%%%%%%%%%%%%%%%%%%%%%%%%%%%%%%%%%%%%%%%%%%%%%%%%%%%%%%%%%%%%%%%%%%%%%
%%%%%%%%%%%%%%%%%%%%%%%%%%%%%%%%%%%%%%%%%%%%%%%%%%%%%%%%%%%%%%%%%%%%%%%%%%%%%%%%%%%%%%%%%%%%%%%%%%%
%%%%%%%%%%%%%%%%%%%%%%%%%%%%%%%%%%%%%%%%%%%%%%%%%%%%%%%%%%%%%%%%%%%%%%%%%%%%%%%%%%%%%%%%%%%%%%%%%%%
\section{Introduction}

Gravity has become the stumbling block in our search for a unified theory, which probably will lead to an understanding of the origin of our universe to the late time cosmic acceleration. On the other hand, even though the standard model of strong and electroweak interactions can explain a vast landscape of experimental results, it continues to have some longstanding unresolved issues, which strongly suggests to look for physics beyond the standard model. One of the major drawback of the standard model is the necessity of fine tuning, which originates from the large hierarchy between the electroweak and the Planck scale, known as the gauge hierarchy problem. It is remarkable that gravity provides a very novel solution to this fine tuning problem through the existence of extra dimensions. Such a gravity based resolution of the gauge hierarchy problem was elegantly described by Randall and Sundrum, where a single extra dimension with manifold structure $S^{1}/Z_{2}$ was 
assumed, resulting in two branes (hypersurfaces of $(3+1)$ dimensions) located at orbifold fixed points with positive and negative tensions. Subsequently, starting from the Einstein's equations in the bulk (higher dimensional spacetime) with a negative cosmological constant, they could show that physical mass of a field confined on the negative tension brane is in the weak scale, due to an exponential suppression, whose origin traces back to gravity. There have been numerous works later on to clarify some of the disadvantages of this model, as well as in generalizations to more complex settings (for a representative class of works see \cite{Randall:1999ee,Verlinde:1999fy,Djouadi:2005gi,Garriga:1999yh,Dadhich:2000am,Csaki:1999jh,Chacko:1999eb,Cohen:1999ia,
Antoniadis:1998ig,Rubakov:1983bb,Chung:1999zs,Cline:1999ts,Nihei:1999mt}). One of the key feature of the Randall-Sundrum model is the appearance of an additional four dimensional massless scalar field having no dynamics. This is an undesirable feature, since without a stabilization mechanism one cannot arrive at the desired exponential suppression.

Unfortunately, gravity sector alone could not cure this problem. One had to introduce an additional scalar field in the bulk, whose action when integrated over the extra spatial dimension, provided the potential necessary for stabilization. Any fluctuation about this stabilized value leads to a scalar degree of freedom, called the radion field. There have been numerous studies later on, regarding the details of the stabilization mechanism, corresponding collider signatures and of course, various generalizations, e.g., time dependent stabilization of the radion field (for a small sample of works see \cite{Goldberger:1999uk,Goldberger:1999un,Lipatov:2016ayn,Das:2016dtz,Arun:2016csq,Bazeia:2014dea,Cox:2013rva,Anand:2014vqa,
Chakraborty:2015zxc,Chakraborty:2013ipa}, and also the references therein).  

Even though the above scheme of solving the gauge hierarchy problem looked promising, but there is one aspect, namely the introduction of a bulk scalar which is put in by hand. It would be really intriguing if the whole solution, i.e., the exponential warping as well as the stabilization can come from gravity alone. Further excitement will follow if the radion field so obtained has observable consequences at the collider experiments. In this work, we explore the above possibility and demonstrate explicitly that one can have the desired warping as well as can stabilize the radion field using \emph{only} gravitational interactions! This is achieved by introducing higher curvature corrections to the \EH action, which is expected, since the bulk spacetime is governed by Planck scale physics. We further delineate on the phenomenology of the associated radion field, whose potential is being supplied by the higher curvature corrections and demonstrate the significance for collider physics. The 
phenomenological study enables one to probe the properties of gravitational physics, in particular that of higher curvature gravity, using colliders, leading to new avenues of exploration. 

We have organized the paper as follows: We start with a brief introduction to higher curvature gravity and the particular model we will be interested in. Proceeding further we demonstrate how one can have both the exponential warping and radion stabilization in this scenario, the main theme of this work. Finally we discuss the radion phenomenology and comment on possible collider signatures of our model before pointing out future directions of exploration.
%%%%%%%%%%%%%%%%%%%%%%%%%%%%%%%%%%%%%%%%%%%%%%%%%%%%%%%%%%%%%%%%%%%%%%%%%%%%%%%%%%%%%%%%%%%%%%%%%%%
%%%%%%%%%%%%%%%%%%%%%%%%%%%%%%%%%%%%%%%%%%%%%%%%%%%%%%%%%%%%%%%%%%%%%%%%%%%%%%%%%%%%%%%%%%%%%%%%%%%
%%%%%%%%%%%%%%%%%%%%%%%%%%%%%%%%%%%%%%%%%%%%%%%%%%%%%%%%%%%%%%%%%%%%%%%%%%%%%%%%%%%%%%%%%%%%%%%%%%%
\section{Background: Higher curvature gravity}

It is generally believed that at high energies (or, small length scales) the \EH action must be supplemented with higher curvature corrections respecting the diffeomorphism invariance of the action. There are several possibilities for the same, two such candidates being $f(R)$ theories of gravity and Lanczos-Lovelock models of gravity. The Lanczos-Lovelock models are more complicated, but is free of ghosts due to its quasi-linear structure \cite{Chakraborty:2015kva,Chakraborty:2015wma,Chakraborty:2014joa,Chakraborty:2014rga,Dadhich:2008df,Padmanabhan:2013xyr}. On the other hand the $f(R)$ models need special care and must satisfy few conditions in order to ensure its ghost free behavior. The success of $f(R)$ models lie in its excellent match with observations as far as the cosmological arena is considered. Further, $f(R)$ models with a certain constraint on its parameters can also evade the solar system tests as well \cite{Pogosian:2007sw,
Capozziello:2005bu,Capozziello:2007ms,Sotiriou:2005xe,Capozziello:2006jj,Nojiri:2001ae,Nojiri:2010wj,
Sotiriou:2008rp,DeFelice:2010aj,Chakraborty:2014xla,Chakraborty:2015bja,Chakraborty:2015taq,Nojiri:2003ft,Nojiri:2007as,Hu:2007nk}. In this work, it will be sufficient for our purpose to focus on the $f(R)$ theories of gravity, satisfying a couple of constraints ensuring its ghost free behavior. 

We will work with a five dimensional spacetime consisting of a single extra spacelike coordinate $y$. The extra dimension will assumed to be compact with a $S^{1}/Z_{2}$ orbifold structure. Alike the Randall-Sundrum scenario, two branes are located at the orbifold fixed points $y=0,\pi$ respectively, with $y$ and $-y$ identified. The bulk gravitational action is assumed to be of the following form
%%%%%%%%%%%%%%%%%%%%%%%%%%%%%%%%%%%%%%%%%%%%%%%%%%%%%%%%%%%%%%%%%%%%%%%%%%
\begin{align}\label{Eq_01}
\mathcal{A}=\int d^{4}xdy~\sqrt{-g}\left[\frac{1}{2\kappa _{5}^{2}}\left\lbrace f(R)\right\rbrace -\Lambda\right]=\int d^{4}xdy~\sqrt{-g}\left[\frac{1}{2\kappa _{5}^{2}}\left\lbrace R+\alpha R^{2}-|\beta| R^{4}\right\rbrace -\Lambda\right]~,
\end{align}
%%%%%%%%%%%%%%%%%%%%%%%%%%%%%%%%%%%%%%%%%%%%%%%%%%%%%%%%%%%%%%%%%%%%%%%%%%
where, $\kappa _{5}$ is the five dimensional gravitational constant with mass dimension $-3/2$, $\Lambda$ being the negative bulk cosmological constant with mass dimension $5$, $\alpha$ and $\beta$ are the higher curvature couplings having mass dimensions of $-2$ and $-6$ respectively. The structure of the above Lagrangian has been inferred from the ghost free criterion, which reads, $f'(R)>0$, $f'(R)<1$ and $f''(R)>0$ \cite{Pogosian:2007sw}. For convenience we will switch to units where Planck mass has been set to unity, using which the above conditions lead to $\alpha >0$ as well as $\alpha >|\beta|$. Note that the model must satisfy these criterion at all curvature scales. An exact warped geometric solution to the above gravitational action has been derived recently in \cite{Chakraborty:2016ydo}, which reads
%%%%%%%%%%%%%%%%%%%%%%%%%%%%%%%%%%%%%%%%%%%%%%%%%%%%%%%%%%%%%%%%%%%%%%%%%%
\begin{align}\label{Eq_02}
ds^{2}=f(y)\left[e^{-2A(y)}\eta _{\mu \nu}dx^{\mu}dx^{\nu}+r_{c}^{2}dy^{2}\right];\qquad A(y)=kr_{c}y+\frac{\kappa _{5}^{2}v^{2}}{12}\exp \left(-\frac{4b_{0}}{\kappa _{5} ^{2}}r_{c}y \right)~.
\end{align}
%%%%%%%%%%%%%%%%%%%%%%%%%%%%%%%%%%%%%%%%%%%%%%%%%%%%%%%%%%%%%%%%%%%%%%%%%%
Here $v$ is a constant of mass dimension $3/2$, we have also defined $b_{0}=(9\kappa _{5}^{2}\sqrt{|\beta|}/32\alpha ^{2})$ and $k=\sqrt{-\Lambda \kappa _{5}^{2}/6}$ for convenience. For completeness we also present the form of the function $f(y)$ appearing in \ref{Eq_02}, having the following structure
%%%%%%%%%%%%%%%%%%%%%%%%%%%%%%%%%%%%%%%%%%%%%%%%%%%%%%%%%%%%%%%%%%%%%%%%%%
\begin{align}\label{Eq_03}
f(y)=\left[1+\frac{\sqrt{3}\kappa _{5}v}{2}\exp\left(-\frac{2b_{0}}{\kappa _{5}^{2}}r_{c}y\right)
-\frac{3\sqrt{3}|\beta|\kappa _{5}^{3}v^{3}}{16\alpha ^{3}}\exp\left(-\frac{6b_{0}}{\kappa _{5}^{2}}r_{c}y\right)\right]^{-2/3}~.
\end{align}
%%%%%%%%%%%%%%%%%%%%%%%%%%%%%%%%%%%%%%%%%%%%%%%%%%%%%%%%%%%%%%%%%%%%%%%%%%
Note that the solution derived above has the desired exponential warping, through the term $kr_{c}y$ in the warp factor. Such that on the visible brane at $y=\pi$, the physical mass of any field will be suppressed by $\exp(-kr_{c}\pi)$, leading to weak scale behavior. This is alike the Randall-Sundrum scenario, where introducing gravity alone leads to the desired weak scale phenomenology with a choice of $kr_{c}\simeq 12$. We would like to emphasize the fact that the effect of higher curvature gravity is through some combinations of $\alpha$ and $\beta$ both, keeping only $\alpha$ or $\beta$ is not sufficient to get the desired warping. The next hurdle is to provide a stabilization mechanism for the radion, which will drive the radion field to its value $r_{c}$, compatible with the exponential suppression. This is what we will elaborate on in the next section.
%%%%%%%%%%%%%%%%%%%%%%%%%%%%%%%%%%%%%%%%%%%%%%%%%%%%%%%%%%%%%%%%%%%%%%%%%%%%%%%%%%%%%%%%%%%%%%%%%%%
%%%%%%%%%%%%%%%%%%%%%%%%%%%%%%%%%%%%%%%%%%%%%%%%%%%%%%%%%%%%%%%%%%%%%%%%%%%%%%%%%%%%%%%%%%%%%%%%%%%
%%%%%%%%%%%%%%%%%%%%%%%%%%%%%%%%%%%%%%%%%%%%%%%%%%%%%%%%%%%%%%%%%%%%%%%%%%%%%%%%%%%%%%%%%%%%%%%%%%%
\section{Stabilizing the radion using higher curvature gravity}\label{Sec_03}

In the original Randall-Sundrum scenario, one needs to introduce a bulk scalar field in order to achieve the stabilization of the radion field. The reason being, the Lagrangian for the Randall-Sundrum scenario, which is the Ricci scalar, has no potential term for the radion field. The bulk field introduces such a potential, and thereby satisfies both the suppression of the Planck scale and the stabilization of the radion field. We will show that, since our action in \ref{Eq_01} has higher curvature terms, if one evaluates the same for the metric given in \ref{Eq_02} and \ref{Eq_03}, it will \emph{naturally} lead to a potential for the radion field and hence one can stabilize the same \emph{without} ever introducing a bulk scalar field. The additional degree of freedom originating from the higher curvature terms actually plays the role of a stabilizing field. As a consequence, the stabilized value would depend upon the parameters $\alpha$ and $\beta$ appearing as the couplings of the higher curvature terms. It will turn out that the condition $kr_{c}\simeq 12$ is essentially a condition on these higher curvature couplings and the bulk cosmological constant. 

As we have already laid out the principles involved, we will proceed directly to the computation and will work exclusively with the higher curvature action presented in \ref{Eq_01}. Given the metric ansatz in \ref{Eq_02} one can evaluate the Ricci scalar in a straightforward manner. Then one has to substitute the Ricci scalar in the gravitational part of the action and obtain the corresponding $f(R)$ Lagrangian, which to leading orders in the coupling parameters reads (for details see \ref{App_01}),
%%%%%%%%%%%%%%%%%%%%%%%%%%%%%%%%%%%%%%%%%%%%%%%%%%%%%%%%%%%%%%%%%%%%%%%%
\begin{align}
f(R)\simeq -20k^{2}-\frac{20}{\sqrt{3}}k^{2}\left(\kappa _{5}v\right)e^{-2b_{0}r_{c}y/\kappa _{5}^{2}}+k\left(\kappa _{5}^{2}v^{2}\right)\frac{15\sqrt{|\beta|}}{4\alpha ^{2}} e^{-4b_{0}r_{c}y/\kappa _{5}^{2}}~.
\end{align}
%%%%%%%%%%%%%%%%%%%%%%%%%%%%%%%%%%%%%%%%%%%%%%%%%%%%%%%%%%%%%%%%%%%%%%%%
Given this form of the $f(R)$ Lagrangian, one can substitute the same in the bulk action, i.e., \ref{Eq_01} and then integrate out the extra dimensional coordinate $y$ over the interval $[0,\pi]$, thanks to the orbifold symmetry. In this integration it will turn out that the contribution coming from the lower limit $y=0$ is independent of the radion field $r_{c}$ and hence adds a constant contribution to the radion potential, while the contribution from $y=\pi$ does have dependence on the radion field and shall serve as the radion potential. Introducing a new field $\phi(r_{c})=\Phi \exp[- kr_{c}\pi]$, where $\Phi =\sqrt{24/k\kappa _{5}^{2}}$, we finally obtain the potential for the radion field (or, equivalently for the new field $\phi$) to yield,
%%%%%%%%%%%%%%%%%%%%%%%%%%%%%%%%%%%%%%%%%%%%%%%%%%%%%%%%%%%%%%%%%%%%%%%%%%
\begin{align}\label{Eq_04}
V(\phi)&=\int dy~\left[\frac{f(R)}{2\kappa _{5}^{2}}-\Lambda\right] 
\nonumber
\\
&\simeq \frac{5k}{2\kappa _{5}^{2}}\left(\frac{\phi}{\Phi}\right)^{4}\Bigg[-1+\frac{\sqrt{3}\kappa _{5}v}{2}\left(\frac{\phi}{\Phi}\right)^{\delta}+\frac{9\kappa _{5}^{2}v^{2}\sqrt{|\beta|}}{48\alpha ^{2}}\left(\frac{\phi}{\Phi}\right)^{2\delta}\Bigg]
+\textrm{constant}
\end{align}
%%%%%%%%%%%%%%%%%%%%%%%%%%%%%%%%%%%%%%%%%%%%%%%%%%%%%%%%%%%%%%%%%%%%%%%%%%
where, $\delta =(2b_{0}/k \kappa _{5}^{2})=(9\sqrt{|\beta|}/16k\alpha ^{2})$, is a dimensionless constant. The usefulness of this quantity $\phi$ follows from the fact that one can upgrade it immediately to the status of a four dimensional scalar field with $r_{c}\rightarrow r(x)$, spacetime dependent brane separation. Such that the vacuum expectation value of the same is given by $r_{c}$ and thus $\phi(r_{c})$ will denote the vacuum expectation value of $\phi$ \cite{Goldberger:1999uk,Goldberger:1999un}. Note that the field $\phi$ is merely a constant depending on the radion field. If the field is being upgraded to depend on the spacetime coordinates, then the potential structure will remain identical, however it will inherit the canonical kinetic term in the action as well (see, for example \cite{Goldberger:1999un}). 

Note that the above potential is very much similar to the one obtained in \cite{Goldberger:1999uk,Goldberger:1999un} using bulk scalar field, with $\delta$ identified as $(m^{2}/4k^{2})$, where $m$ is the mass of the bulk scalar. Thus the higher curvature terms act as a source for the radion mass as we will explicitly illustrate later. Further in the above scheme the condition $\delta <1$ is identically satisfied, since the couplings to higher and higher curvature terms are more and more suppressed. Thus the scenario with higher curvature gravity leads to an identical situation as that of introducing a bulk scalar field, but follows from the gravity sector alone. The stabilized value of the radion field can be obtained by finding out the minima of the potential in \ref{Eq_04}, which is a solution of the equation $\partial V/\partial \phi =0$, leading to the following expression for $r_{c}$ (see \ref{App_Eq01} in \ref{App_01}),
%%%%%%%%%%%%%%%%%%%%%%%%%%%%%%%%%%%%%%%%%%%%%%%%%%%%%%%%%%%%%%%%%%%%%%%%%%
\begin{align}\label{Eq_05}
kr_{c}=\frac{16\alpha ^{2}}{9 \pi \sqrt{|\beta|}}\sqrt{-\frac{\Lambda \kappa _{5}^{2}}{6}}~
\ln \left[\frac{\frac{\sqrt{3}\kappa _{5}v\sqrt{|\beta|}}{4k\alpha ^{2}}}{\sqrt{1+\frac{\sqrt{3}\kappa _{5}v\sqrt{|\beta|}}{2k\alpha ^{2}}}-1}\right]
\end{align}
%%%%%%%%%%%%%%%%%%%%%%%%%%%%%%%%%%%%%%%%%%%%%%%%%%%%%%%%%%%%%%%%%%%%%%%%%%
Thus with $\kappa _{5}v \sim 40$ and $\sqrt{|\beta|}/k\alpha ^{2}\simeq 1/20$, the logarithmic term becomes of order unity and then one readily obtains $kr_{c}\simeq 12$, the value desired for exponential warping. Hence starting from a pure gravitational action, with higher curvature corrections, one can produce an exponential warping as well as a proper stabilized value for the brane separation without ever introducing any additional structure. Further the desired warping to address the hierarchy problem leads to a relation between the higher curvature couplings and the bulk cosmological constant. This completes what we set out to prove, i.e., derivation of the exponential warping leading to weak scale physics and a proper stabilization mechanism for the brane separation, both from the gravitational dynamics alone. 

For the sake of completeness we would like to discuss on the choice of $f(R)$ gravity and its role in radion stabilization. Since we are working in the higher curvature regime, where bulk effects are important, it makes sense to add terms like $R^{n}$ to the \EH action, with $n$ positive. The first such choice corresponds to adding a $R^{2}$ term, which would lead to the original Goldberger-Wise scenario in the scalar tensor representation. The next leading order term in a bulk spacetime with negative cosmological constant, free of ghosts correspond to the one presented in \ref{Eq_01}. Interestingly for this situation one can solve for the full scalar coupled Einstein's equations as depicted in \cite{Chakraborty:2016ydo} and the situation will differ considerably from the Goldberger-Wise scenario. In principle one can add more higher order terms to the Lagrangian, however in those scenarios one would not be able to solve the full back-reacted problem in scalar-tensor representation. Moreover, these terms will be further suppressed and will contribute insignificant corrections over and above \ref{Eq_05}. Thus the scenario presented in this work captures all the essential features and is simple enough to be solved in an exact manner. This motivates the choice presented in \ref{Eq_01}. Given the above, it will be worthwhile to spend some time discussing the corresponding scenario in the scalar-tensor representation \cite{Chakraborty:2016ydo}, which will bring out the difference of our approach with the existing ones. This is what we elaborate in the next section.
%%%%%%%%%%%%%%%%%%%%%%%%%%%%%%%%%%%%%%%%%%%%%%%%%%%%%%%%%%%%%%%%%%%%%%%%%%%%%%%%%%%%%%%%%%%%%%%%%%%
%%%%%%%%%%%%%%%%%%%%%%%%%%%%%%%%%%%%%%%%%%%%%%%%%%%%%%%%%%%%%%%%%%%%%%%%%%%%%%%%%%%%%%%%%%%%%%%%%%%
%%%%%%%%%%%%%%%%%%%%%%%%%%%%%%%%%%%%%%%%%%%%%%%%%%%%%%%%%%%%%%%%%%%%%%%%%%%%%%%%%%%%%%%%%%%%%%%%%%%
\section{Stabilization in the Einstein frame: Scalar-tensor representation}

It is well known that any $f(R)$ gravity model is mathematically equivalent to a dual scalar-tensor representation \cite{Barrow:1988xh,Capozziello:1996xg,Nojiri:2010wj,Sotiriou:2008rp,DeFelice:2010aj,Anand:2014vqa,Bahamonde:2016wmz,
Parry:2005eb,Catena:2006bd,Chiba:2013mha,Bhattacharya:2016lup}. The mathematical equivalence follows from the transformation of the Jordan frame action to the Einstein frame aka conformal transformation. Surprisingly, this equivalence holds in lower dimensions as well, viz., if one starts from a higher dimensional action and projects on to a lower dimensional hypersurface the field equations are still connected by conformal transformation, provided one exercise caution about the boundary contributions. Despite the mathematical equivalence, there are situations where the two scenarios are not \emph{physically} equivalent, e.g., in cosmological scenarios the $f(R)$ frame may lead to late time acceleration, while the Einstein frame advocates late time deceleration \cite{Briscese:2006xu,Bahamonde:2016wmz,Capozziello:2010sc}. The issue of physical inequivalence becomes important if the spacetime inherits a singularity or is undergoing a quick evolution phase. In particular, using reconstruction technique \cite{Nojiri:2009kx,Nojiri:2006be,Nojiri:2009xh,Carloni:2010ph} it is possible to generate $f(R)$ theories explaining the early inflationary phase to late time accelerating phase of the universe, all of which ultimately results into finite time future singularity. The existence of future singularity often breaks the equivalence with scalar-tensor representation \cite{Bamba:2008ut,Bahamonde:2016wmz}. However in absence of singularity \cite{Antoniadis:1993jc} equivalence of $f(R)$ theories with scalar tensor theories does exist. The situation discussed in this work has no such singularity in the spacetime structure, as evident from \ref{Eq_02} and \ref{Eq_03} respectively. Thus one may safely use the equivalence between $f(R)$ and scalar tensor theories. 

Given this input, it will be worthwhile to explore the corresponding situation in this context, namely how the radion stabilization is affected as one considers the dual picture in the Einstein frame and contrast the same with the stabilization already discussed in \ref{Sec_03}. For the $f(R)$ action under our consideration, the corresponding action in the Einstein frame becomes,
%%%%%%%%%%%%%%%%%%%%%%%%%%%%%%%%%%%%%%%%%%%%%%%%%%%%%%%%%%%%%%%%%%%%%%%%
\begin{align}
\mathcal{A}=\int d^{5}x\sqrt{-g}\left[\frac{R}{2\kappa _{5}^{2}}-\frac{1}{2}g^{ab}\nabla _{a}\psi \nabla _{b}\psi -V(\psi)-\Lambda \right]
\label{ST_01}
\end{align}
%%%%%%%%%%%%%%%%%%%%%%%%%%%%%%%%%%%%%%%%%%%%%%%%%%%%%%%%%%%%%%%%%%%%%%%%
where $\psi$ is the scalar field in the dual picture defined as, $\kappa _{5}\psi=(2/\sqrt{3})\ln (1+f')$ and the corresponding potential becomes
%%%%%%%%%%%%%%%%%%%%%%%%%%%%%%%%%%%%%%%%%%%%%%%%%%%%%%%%%%%%%%%%%%%%%%%%%
\begin{align}
V(\psi)=\frac{3}{32\alpha}\psi ^{2}-\frac{\kappa _{5}^{2}}{6}k^{2}\delta ^{2}\psi ^{4}
\end{align}
%%%%%%%%%%%%%%%%%%%%%%%%%%%%%%%%%%%%%%%%%%%%%%%%%%%%%%%%%%%%%%%%%%%%%%%%%%
The expression for the potential brings out the key difference between the Goldberger-Wise stabilization mechanism and the one advocated here --- the potential for the scalar field in Goldberger-Wise mechanism lacks the quartic term present in our analysis. Incidentally, the presence of this quartic term helps to solve the full back-reacted problem, while the original stabilization proposal was without incorporating the back-reaction of the scalar field. Hence the stabilized value of the radion field derived above incorporates the effect of the quartic potential, as well as the back-reaction of the scalar field on the spacetime geometry and differs from the standard scenarios. The above structure of the potential also shows the reason for neglecting further higher curvature terms (e.g., $R^{6}$) in the action \footnote{It is evident that, the higher curvature terms are more and more suppressed, and we choose to work with the first two leading order corrections, compatible with ghost free criterion, to the \EH 
action.}. The scalar tensor representation with such higher curvature terms will involve more complicated potentials and hence cannot be solved in full generality by incorporating the back-reaction as well. Further the $R^{4}$ term in the action leads to the leading order departure from the Goldberger-Wise action, which we have studied in this paper. Additional higher curvature terms would lead to further sub-leading corrections and thus can be neglected.

Given the action in the Einstein frame, one can invoke the corresponding solution (derived in \ref{App_01}) and integrate out the extra dimensional part present in the action. This will result in a potential for the radion field $r_{c}$, alike the Goldberger-Wise mechanism, whose minima would yield the following stabilized value of the radion field,
%%%%%%%%%%%%%%%%%%%%%%%%%%%%%%%%%%%%%%%%%%%%%%%%%%%%%%%%%%%%%%%%%%%%%%%%%%%%%
\begin{align}
kr_{c}\simeq \frac{16\alpha ^{2}}{9 \pi \sqrt{|\beta|}}\sqrt{-\frac{\Lambda \kappa _{5}^{2}}{6}}~\ln \left(\frac{3\sqrt{|\beta|}\kappa _{5}^{2}\psi _{0}^{2}}{64k\alpha ^{2}} \right)
\end{align}
%%%%%%%%%%%%%%%%%%%%%%%%%%%%%%%%%%%%%%%%%%%%%%%%%%%%%%%%%%%%%%%%%%%%%%%%%%%%%%
where $\psi _{0}$ is the value of the dual scalar field in the $y=0$ brane. Comparison with \ref{Eq_05} reveals that the leading order behavior (i.e., the term outside logarithm) of the stabilized value of the radion field is identical in both Jordan and Einstein frame. This observation explicitly demonstrates that, in both these frames the radion is stabilized to the desired value necessary for exponential suppression of the Planck scale. Thus unlike various scenarios with either singularity or a quick evolution  (where the two frames are physically inequivalent) in this particular situation the physical equivalence between the two frames is manifest. 

At this stage, we may point out another alternative possibility of stabilizing the radion field, by incorporating quantum effects of the bulk scalar field at nonzero temperature \cite{Brevik:2000vt,Brevik:2001bi}. In particular, one can think of the resulting thermal fluctuations to generate a modulus potential which may inherit a minimum, thereby stabilizing the brane separation. The brane separation necessary to solve the hierarchy problem involves considering a low temperature limit of the free energy associated with the bulk quantum field, which may have connections with the AdS/CFT correspondence \cite{Brevik:2000vt,Brevik:2001bi}. However, in this context as well, one generally neglects the effect of back-reaction and treats the bulk quantum field to be siting on the fixed Anti-de Sitter background, unlike the scenario we have depicted. Thus after explicitly establishing the differences between our approach and the existing ones, we now concentrate on the phenomenology of the radion field, viz. mass of the radion field and its interaction with the standard model fields in the next section.
%%%%%%%%%%%%%%%%%%%%%%%%%%%%%%%%%%%%%%%%%%%%%%%%%%%%%%%%%%%%%%%%%%%%%%%%%%%%%%%%%%%%%%%%%%%%%%%%%%%
%%%%%%%%%%%%%%%%%%%%%%%%%%%%%%%%%%%%%%%%%%%%%%%%%%%%%%%%%%%%%%%%%%%%%%%%%%%%%%%%%%%%%%%%%%%%%%%%%%%
%%%%%%%%%%%%%%%%%%%%%%%%%%%%%%%%%%%%%%%%%%%%%%%%%%%%%%%%%%%%%%%%%%%%%%%%%%%%%%%%%%%%%%%%%%%%%%%%%%%
\section{Some applications of the stabilized radion field}

In this section we will briefly discuss two possible applications of the stabilized radion field in two diverse physical contexts. The first application will discuss the implications of this stabilized radion from the perspective of particle physics, while the other will address the effect of the radion field on the inflationary paradigm. In both these contexts we will demonstrate that the stabilization of the radion field (which has its origin in higher curvature gravity) plays a crucial role and leads to interesting additional structure in the corresponding situations having possible observational consequences.
%%%%%%%%%%%%%%%%%%%%%%%%%%%%%%%%%%%%%%%%%%%%%%%%%%%%%%%%%%%%%%%%%%%%%
%%%%%%%%%%%%%%%%%%%%%%%%%%%%%%%%%%%%%%%%%%%%%%%%%%%%%%%%%%%%%%%%%%%%%
%%%%%%%%%%%%%%%%%%%%%%%%%%%%%%%%%%%%%%%%%%%%%%%%%%%%%%%%%%%%%%%%%%%%%
\subsection{Phenomenology of the stabilized radion}

Given the potential $V(\phi)$, one can immediately obtain the mass of the $\phi$ excitation by computing $\partial ^{2}V/\partial \phi ^{2}$ and then expressing the same using $r_{c}$, given in \ref{Eq_05}. Performing the same, one arrives at the following expression for radion mass,
%%%%%%%%%%%%%%%%%%%%%%%%%%%%%%%%%%%%%%%%%%%%%%%%%%%%%%%%%%%%%%%%%%%%%%%%%%
\begin{align}\label{Eq_06}
m_{\phi}^{2}\equiv \frac{\partial ^{2}V}{\partial \phi ^{2}}(r_{c})=\frac{5k^{2}\kappa _{5}^{2}v^{2}}{18}
\left(\frac{\phi(r_{c})}{\Phi} \right)^{2\delta}\delta ^{2}e^{-2kr_{c}\pi}~,
\end{align}
%%%%%%%%%%%%%%%%%%%%%%%%%%%%%%%%%%%%%%%%%%%%%%%%%%%%%%%%%%%%%%%%%%%%%%%%%%
where, $\phi(r_{c})$ can be obtained from \ref{Eq_05}. It is evident from the expression for $m_{\phi}$, that the there is an exponential suppression of the radion mass, which leads to a weak scale value from a Planck scale quantity. Note that this expression is very much similar to the result obtained in \cite{Goldberger:1999uk,Goldberger:1999un}, with the identification of $\delta$ with $m^{2}/k^{2}$. But with one difference, which is caused by the $(\phi(r_{c})/\Phi)^{2\delta}$ term in \ref{Eq_06}. This results in a decrease in the radion mass as compared to the Goldberger-Wise scenario described in \cite{Goldberger:1999un}. However at the same time the choice of $\kappa _{5}v$ also becomes important. To see the difference in a quantitative manner, consider the following situation: $\kappa _{5}v\sim 40, \delta \sim 1/32$, using which one obtains $kr_{c}\sim 12$, leading to, $m_{\phi}\sim 0.02 k^{2}e^{-2kr_{c}\pi}$. While for the standard Goldberger-Wise scenario one would have obtained $m_{
\phi}^{2}\sim 0.05k^{2}e^{-2kr_{c}\pi}$. This explicitly shows that the radion mass in our approach is lighter (in this case two times) compared to the one obtained in \cite{Goldberger:1999un}, depicting quantitatively the difference between these two approaches. Further note that the results in \cite{Goldberger:1999uk,Goldberger:1999un} were derived in the context of a bulk scalar field, here we derive the same but from a \emph{purely} gravitational standpoint. Similar exponential suppression will drive the masses of the low lying Kaluza-Klein excitations to TeV scale \cite{DeWolfe:1999cp,Mirabelli:1998rt,Goldberger:1999wh,Dudas:2005vna,Hewett:1998sn,Das:2011fb,Davoudiasl:1999tf,Davoudiasl:2000wi,
Chakraborty:2014xda,Chakraborty:2014zya,Hewett:2016omf,Giddings:2016sfr,Oliveira:2014kla,Cho:2013mva}. In the context of radion mass however there is one further suppressing factor, namely $\delta ^{2}$ and thus radion mass will be a bit smaller compared to the low lying Kaluza-Klein excitations of the bulk field. Since the radion mass, as presented in this work is completely of higher curvature origin, this suggests that detection of the radion field $\phi$ may be a hint not only of higher dimensions but also of higher curvature gravity.   

In order to find out coupling of the radion field to standard model particles, note that in this scenario radion appears as a gravitational degree of freedom and since gravity couples to all kinds of matter, so will the radion. Since the standard model fields are confined on the brane hypersurfaces, it is clear that they will couple to the induced metric on the brane. On the brane located at $y=0$, the warp factor becomes unity and the induced metric is proportional to $\eta _{\mu \nu}$. Thus fields confined to $y=0$ hypersurface does not couple to the radion field. On the other hand, on the visible brane (the hypersurface $y=\pi$) the induced metric is $(\phi/\Phi)^{2}\eta _{\mu \nu}$ and thus radion field will couple to the standard model fields. Here, $\phi=\phi(r_{c})+\delta \phi$, where $\delta \phi$ is the fluctuations around the minimum value $\phi(r_{c})$. In the case of a scalar field, say the standard model Higgs $h(x)$, one can reabsorb the factors of $\Phi$ by redefining $h\rightarrow (\Phi/\phi(
r_{c}))h(x)$, such that physical Higgs mass becomes $m_{0}e^{-kr_{c}\pi}$. For a Planck 
scale bare mass $m_{0}$ and $kr_{c}\simeq 12$ one immediately obtains the physical mass in the weak scale. Further the corresponding interaction term of a standard model field with the radion will involve 
%%%%%%%%%%%%%%%%%%%%%%%%%%%%%%%%%%%%%%%%%%%%%%%%%%%%%%%%%%%%%%%%%%%%%%%%%%
\begin{align}
L_{\rm int}=\frac{1}{\phi(r_{c})}\delta \phi T^{\mu}_{\mu}\equiv \frac{1}{\Lambda _{\phi}}\delta \phi T^{\mu}_{\mu}~,
\end{align}
%%%%%%%%%%%%%%%%%%%%%%%%%%%%%%%%%%%%%%%%%%%%%%%%%%%%%%%%%%%%%%%%%%%%%%%%%%
where $T^{\mu}_{\mu}$ stands for the trace of the energy momentum tensor of the standard model field and $\Lambda _{\phi}$ defines the coupling of the radion to the standard model fields. (Since radion is a gravitational degree of freedom it has to couple to some combination of the matter energy momentum tensor). Note that for large $\Lambda _{\Phi}$, the coupling becomes small and as a consequence radion will couple weakly to the standard model fields. In particular recent bounds on both radion mass and radion coupling strength shows that they are not independent, if radion mass is smaller the coupling will be larger and vice versa. To get a numerical estimate, one must provide an estimate for the value of $k$ in Planck units. For $k\sim 0.1$, the coupling satisfies the following stringent bound $\Lambda _{\phi}>14.3~\textrm{TeV}$ \cite{Cho:2013mva,CMS:ril,Davoudiasl:2012xd,Kubota:2012in,Frank:2011kz,Mahanta:2000ci}. This leads to very weak coupling between radion and standard model fields, 
resulting in non-detectability of the radion field (even though the mass can be as low 
as $200 \textrm{GeV}$). At the same time for $k\sim 1$, in Planck units, the coupling can become $\Lambda _{\phi}\sim 5~\textrm{TeV}$. Even though in this case the coupling is not very weak, but the radion mass becomes higher $\sim 1 \textrm{TeV}$. Thus coupling and mass of the radion field are inversely related, making it difficult to detect in the current generation collider experiments. 

For completeness, let us discuss the implications of our model in the context of electroweak precession measurements \cite{Csaki:2000zn}. As emphasized earlier, the radion will linearly couple to standard model fields, e.g., the gauge bosons $W^{\pm}$ and $Z$ with a coupling $\Lambda _{\phi}^{-1}$, which is $(\textrm{TeV})^{-1}$. These interaction terms will produce three point as well as four point interactions between radion and the gauge bosons due to gauge fixing terms as well as higher loop effects in the effective Lagrangian. Fixing the mass of the standard model Higgs boson at $125~\textrm{GeV}$ \cite{CMS:ril}, affects the electroweak parameters such that they become insensitive to the radion mass, since the radion mass is further suppressed by its own vacuum expectation value \cite{Csaki:2000zn}. Despite above, the fact that the mass of the radion field as well as its coupling with standard model fields are in the TeV scale makes the phenomenology of the radion field a nice testbed for higher dimensional as well as higher curvature physics in the next generation colliders.
%%%%%%%%%%%%%%%%%%%%%%%%%%%%%%%%%%%%%%%%%%%%%%%%%%%%%%%%%%%%%%%%%%%%%
%%%%%%%%%%%%%%%%%%%%%%%%%%%%%%%%%%%%%%%%%%%%%%%%%%%%%%%%%%%%%%%%%%%%%
%%%%%%%%%%%%%%%%%%%%%%%%%%%%%%%%%%%%%%%%%%%%%%%%%%%%%%%%%%%%%%%%%%%%%
\subsection{Cosmology and radion stabilization}

Having already discussed the imprints of the stabilization mechanism of the radion field on the phenomenological side, we will presently address the corresponding situation in a cosmological setting. In a more general context, following \cite{Goldberger:1999wh} one should have made the radion field dynamical by considering arbitrary fluctuations around the stabilized value and hence study the dynamics in an arbitrary background. However this will be a complicated exercise, due to presence of higher curvature terms in the gravitational Lagrangian. Thus we concentrate on a specific situation with the radion field depending on time alone, leading to the stabilized value in the cosmological context. Note that this situation has already been analyzed in the context of \gr\ in \cite{Csaki:1999mp} and has been elaborated further in the context of radion stabilization in \cite{Chakraborty:2013ipa}. Thus one may try to find out the cosmology on the brane in presence of higher curvature terms, whose detailed analysis for various cosmological epochs will be presented elsewhere. In this section we will try to answer this question in the context of inflationary paradigm alone. However for completeness, we will also provide some general computations. The metric ansatz suitable for our purpose corresponds to,
%%%%%%%%%%%%%%%%%%%%%%%%%%%%%%%%%%%%%%%%%%%%%%%%%
\begin{align}
ds^{2}=e^{-2A(y)r(t)}\left\{-dt^{2}+a^{2}(t)\left(dx^{2}+dy^{2}+dz^{2}\right) \right\}+r(t)^{2}dy^{2}
\end{align}
%%%%%%%%%%%%%%%%%%%%%%%%%%%%%%%%%%%%%%%%%%%%%%%%%
where, both the branes are located at $y=0$ and $\pi$ respectively and they are assumed to be expanding with the scale factor $a(t)$. Further, the radion field is assumed to be dynamical, such that, $r(t)=r_{c}+\delta r(t)$. Here $r_{c}$ corresponds to the stabilized value of the radion field derived in the earlier sections and $\delta r(t)$ corresponds to the fluctuations around the stabilized value. Further $A(y)$ should take care of all the extra dimensional dependent quantities, which will behave as $ky$ to the leading order. The Ricci scalar derived for the above ansatz, on a $y=\textrm{constant}$ hypersurface becomes,
%%%%%%%%%%%%%%%%%%%%%%%%%%%%%%%%%%%%%%%%%%%%%%%%%
\begin{align}
R=-20k^{2}+6e^{2kry}\left(\frac{\ddot{a}}{a}+\frac{\dot{a}^{2}}{a^{2}} \right)-3\frac{\dot{a}}{a}\frac{\dot{r}}{r}\left(6kry-2\right)e^{2kry}
+e^{2kry}\left\{\frac{\ddot{r}}{r}\left(2-6kry \right)+\frac{\dot{r}^{2}}{r^{2}}kyr\left(6kry-4 \right) \right\}
\end{align}
%%%%%%%%%%%%%%%%%%%%%%%%%%%%%%%%%%%%%%%%%%%%%%%%%
Having obtained the Ricci scalar one can derive the Lagrangian with ease which is given in \ref{Eq_01}. Derivation of the Lagrangian enables one to obtain the corresponding field equations for $a$ and $r$, by varying the scale factor and the radion field respectively. The gravity theory being $f(R)$ in nature, the field equations will definitely inherit higher than second derivatives of the scale factor and the radion field, which reads,
%%%%%%%%%%%%%%%%%%%%%%%%%%%%%%%%%%%%%%%%%%%%%%%%%
\begin{align}
3a^{2}re^{-4kry}f(R)&-6ra^{2}f'(R)\frac{\ddot{a}}{a}+\frac{d^{2}}{dt^{2}}\left\{6ra^{2}e^{-2kry}f'(R) \right\}
-12ra^{2}e^{-2kry}f'(R)\left(\frac{\dot{a}}{a}\right)^{2}
\nonumber
\\
&+3e^{-2kry}\frac{\dot{r}}{r}\frac{\dot{a}}{a}\left(6kry-2\right)ra^{2}f'(R)
+3\frac{d}{dt}\left(e^{-2kry}\frac{\dot{r}}{r}(6kry-2)ra^{2}f'(R) \right)=p
\end{align}
%%%%%%%%%%%%%%%%%%%%%%%%%%%%%%%%%%%%%%%%%%%%%%%%%
In general it looks sufficiently complicated, however in the case of an exponential expansion (with $a(t)=e^{Ht}$), i.e., for inflationary scenario the above equation simplifies a lot. In particular it is possible to approximately solve for the time dependence of the radion field explicitly, which turns out to be decreasing with time, similar with the corresponding situation in \gr\ \cite{Chakraborty:2013ipa}. Thus as the universe expands exponentially, the radion field decreases with time, finally attaining the stabilized value as the inflation ends. Hence the scenario depicted above can also explain a dynamical procedure for stabilization of the radion field, modulo inflationary paradigm. This provides yet another application of the radion field aka higher curvature gravity in the present context.
%%%%%%%%%%%%%%%%%%%%%%%%%%%%%%%%%%%%%%%%%%%%%%%%%%%%%%%%%%%%%%%%%%%%%%%%%%%%%%%%%%%%%%%%%%%%%%%%%%%
%%%%%%%%%%%%%%%%%%%%%%%%%%%%%%%%%%%%%%%%%%%%%%%%%%%%%%%%%%%%%%%%%%%%%%%%%%%%%%%%%%%%%%%%%%%%%%%%%%%
%%%%%%%%%%%%%%%%%%%%%%%%%%%%%%%%%%%%%%%%%%%%%%%%%%%%%%%%%%%%%%%%%%%%%%%%%%%%%%%%%%%%%%%%%%%%%%%%%%%
\section{Discussion}

Gauge hierarchy problem is a very serious fine tuning problem in standard model physics. One avatar of the extra dimensional physics as depicted by Randall and Sundrum has the capability of addressing the gauge hierarchy problem by suppressing the Planck scale to weak scale \emph{gravitationally}. Unfortunately, to stabilize the above scenario one needs to introduce an additional field. In this work, we have shown that the introduction of such a stabilizing field is unnecessary and one can achieve both the suppression to weak scale as well as a stabilization mechanism starting from a higher curvature gravitational action \emph{alone}. In this sense gravity stabilizes itself! 

Starting from a higher curvature gravitational action we have derived an exact warped geometric solution with the extra spatial dimension having $S^{1}/Z_{2}$ orbifold symmetry and two 3-branes located at orbifold fixed points $y=0,\pi$ respectively. We would like to emphasize that the above solution is exact with effects from the higher curvature terms duly accounted for, unlike the original Goldberger-Wise solution where back reaction of the stabilizing field was neglected. In this respect our work is more in tune with \cite{DeWolfe:1999cp}, where also solutions have been derived with inclusion of back-reaction as well. The warp factor again has exponential suppression and thus the Planck scale physics will be reduced to weak scale phenomenon on the visible brane located at $y=\pi$. But, surprisingly the higher curvature terms provide a potential for the radion field as well, whose minima leads to the stabilized value of the radion field \emph{gravitationally}. Further, the stabilized value depends on the 
gravitational 
couplings present in the higher curvature action as well as the bulk cosmological constant, such that a particular choice of these parameters results to $kr_{c}\simeq 12$, leading to the desired warping. Added to the excitement is the result that the radion mass has similar suppression leading to TeV scale physics, which is smaller compared to the low-lying Kaluza-Klein spectrum of the bulk scalar field. Being a gravitational degree of freedom, radion field couples to the standard model fields through the trace of the energy momentum tensor with a coupling having strength $\textrm{TeV}^{-1}$. Once again, we reiterate on the fact that the above analysis has been performed completely in a gravitational physics framework.

The above results open up a broad spectrum of further avenues to explore. From the observational point of view, the above result brings down the dynamics of the higher curvature gravity to a TeV scale phenomenon, which may become accessible in near future collider. In particular the radion mass which depends exclusively on the couplings present in the higher curvature theory may provide the first hint towards observational signatures of higher curvature gravity besides that of higher dimensions. A more careful analysis in this direction can be performed following \cite{Hewett:1998sn,Das:2011fb,Davoudiasl:1999tf,Davoudiasl:2000wi,Chakraborty:2014xda}, where similar analysis for an additional stabilizing scalar field has been carried out. From the theoretical hindsight, it will be worthwhile to understand the phenomenon of generating a potential for the radion field and its subsequent stabilization in the context of Lanczos-Lovelock (or, Einstein-Gauss-Bonnet for simplicity) gravity. It will also be of interest 
to explore the consequences of making the radion field dynamical, in which case the kinetic term will also contribute to the gravitational field equations resulting in distinctive cosmological consequences of this higher curvature brane world scenario, e.g., imprints on the Cosmic Microwave Background, inflationary scenario and so on. 
%%%%%%%%%%%%%%%%%%%%%%%%%%%%%%%%%%%%%%%%%%%%%%%%%%%%%%%%%%%%%%%%%%%%%%%%%%%%%%%%%%%%%%%%%%%%%%%%%%%
%%%%%%%%%%%%%%%%%%%%%%%%%%%%%%%%%%%%%%%%%%%%%%%%%%%%%%%%%%%%%%%%%%%%%%%%%%%%%%%%%%%%%%%%%%%%%%%%%%%
%%%%%%%%%%%%%%%%%%%%%%%%%%%%%%%%%%%%%%%%%%%%%%%%%%%%%%%%%%%%%%%%%%%%%%%%%%%%%%%%%%%%%%%%%%%%%%%%%%%
\section*{Acknowledgement}

Research of S.C. is supported by the SERB-NPDF grant (PDF/2016/001589) from Government of India.

%%%%%%%%%%%%%%%%%%%%%%%%%%%%%%%%%%%%%%%%%%%%%%%%%%%%%%%%%%%%%%%%%%%%%%%%%%%%%%%%%%%%%%%%%%%%%%%%%%%
%%%%%%%%%%%%%%%%%%%%%%%%%%%%%%%%%%%%%%%%%%%%%%%%%%%%%%%%%%%%%%%%%%%%%%%%%%%%%%%%%%%%%%%%%%%%%%%%%%%
%%%%%%%%%%%%%%%%%%%%%%%%%%%%%%%%%%%%%%%%%%%%%%%%%%%%%%%%%%%%%%%%%%%%%%%%%%%%%%%%%%%%%%%%%%%%%%%%%%%
\appendix
\labelformat{section}{Appendix #1} 
\labelformat{subsection}{Appendix #1}
%%%%%%%%%%%%%%%%%%%%%%%%%%%%%%%%%%%%%%%%%%%%%%%%%%%%%%%%%%%%%%%%%%%%%%%%%%%%%%%%%%%%%%%%%%%%%%%%%%%
%%%%%%%%%%%%%%%%%%%%%%%%%%%%%%%%%%%%%%%%%%%%%%%%%%%%%%%%%%%%%%%%%%%%%%%%%%%%%%%%%%%%%%%%%%%%%%%%%%%
%%%%%%%%%%%%%%%%%%%%%%%%%%%%%%%%%%%%%%%%%%%%%%%%%%%%%%%%%%%%%%%%%%%%%%%%%%%%%%%%%%%%%%%%%%%%%%%%%%%
\section{Appendix: Calculational Details}\label{App_01}

In this section, we provide all the calculational details to supplement the results presented in the main text. In order to make the calculation simple, we will introduce the following definitions:
%%%%%%%%%%%%%%%%%%%%%%%%%%%%%%%%%%%%%%%%%%%%%%%%%%%%%%%%%%%%%%%%%%%%%%%%%%
\begin{align}
k=\sqrt{-\frac{\Lambda \kappa _{5}^{2}}{6}};\qquad A_{1}=\frac{\kappa _{5}^{2}v^{2}}{12};\qquad A_{2}=\frac{2b_{0}}{\kappa _{5}^{2}}=\frac{9\sqrt{|\beta|}}{16\alpha ^{2}};\qquad A_{3}=\frac{\sqrt{3}\kappa _{5}v}{2};\qquad A_{4}=-\frac{3\sqrt{3}|\beta|\kappa _{5}^{3}v^{3}}{16\alpha ^{3}}~.
\end{align}
%%%%%%%%%%%%%%%%%%%%%%%%%%%%%%%%%%%%%%%%%%%%%%%%%%%%%%%%%%%%%%%%%%%%%%%%%%
We further divide the appendix in two parts, the first done depicts the situation with $f(R)$ gravity, while the other illustrates the dual scalar-tensor description.
%%%%%%%%%%%%%%%%%%%%%%%%%%%%%%%%%%%%%%%%%%%%%%%%%%%%%%%%%%%%%%%%%%%%%%%%%%%%%%
%%%%%%%%%%%%%%%%%%%%%%%%%%%%%%%%%%%%%%%%%%%%%%%%%%%%%%%%%%%%%%%%%%%%%%%%%%%%%%
%%%%%%%%%%%%%%%%%%%%%%%%%%%%%%%%%%%%%%%%%%%%%%%%%%%%%%%%%%%%%%%%%%%%%%%%%%%%%%
\subsection{Stabilization in f(R) gravity}

The Ricci scalar computed for the metric presented in \ref{Eq_02} turns out to be
%%%%%%%%%%%%%%%%%%%%%%%%%%%%%%%%%%%%%%%%%%%%%%%%%%%%%%%%%%%%%%%%%%%%%%%%%%
\begin{align}
R=\frac{1}{r_{c}^{2}f^{3}(y)}\left[f^{2}(y)\left(8A''-20A'^{2}\right)+4f\left(4A'f'-f'' \right)+f'^{2} \right]~,
\end{align}
%%%%%%%%%%%%%%%%%%%%%%%%%%%%%%%%%%%%%%%%%%%%%%%%%%%%%%%%%%%%%%%%%%%%%%%%%%
where `prime' denotes differentiation with respect to the extra spatial coordinate $y$. Evaluation of the Ricci scalar, given the functions $A(y)$ and $f(y)$ can be performed to leading orders of $\exp(-A_{2}r_{c}y)$ yielding
%%%%%%%%%%%%%%%%%%%%%%%%%%%%%%%%%%%%%%%%%%%%%%%%%%%%%%%%%%%%%%%%%%%%%%%%%%
\begin{align}
R&=-20k^{2}+e^{-A_{2}r_{c}y}\left\lbrace \frac{8}{3}A_{2}^{2}A_{3}+\frac{32}{3}kA_{2}A_{3}-\frac{40}{3}k^{2}A_{3}\right\rbrace
\nonumber
\\
&+e^{-2A_{2}r_{c}y}\Big\lbrace 80kA_{1}A_{2}-\frac{32}{3}kA_{2}A_{3}^{2}+32A_{1}A_{2}^{2}-\frac{20}{3}A_{2}^{2}A_{3}^{2}
+\frac{20}{9}k^{2}A_{3}^{2}
\nonumber
\\
&+\frac{2}{3}A_{3} \left(\frac{8}{3}A_{2}^{2}A_{3}+\frac{32}{3}kA_{2}A_{3}-\frac{40}{3}k^{2}A_{3}\right) \Big\rbrace
\nonumber
\\
&\equiv -20k^{2}+Ee^{-A_{2}r_{c}y}+Fe^{-2A_{2}r_{c}y}~.
\end{align}
%%%%%%%%%%%%%%%%%%%%%%%%%%%%%%%%%%%%%%%%%%%%%%%%%%%%%%%%%%%%%%%%%%%%%%%%%%
Let us now write down the higher curvature Lagrangian of \ref{Eq_01} using the above expression for the Ricci scalar,
%%%%%%%%%%%%%%%%%%%%%%%%%%%%%%%%%%%%%%%%%%%%%%%%%%%%%%%%%%%%%%%%%%%%%%%%%%
\begin{align}
L&=\left(-20k^{2}+Ee^{-A_{2}r_{c}y}+Fe^{-2A_{2}r_{c}y}\right)+\alpha \left(-20k^{2}+Ee^{-A_{2}r_{c}y}+Fe^{-2A_{2}r_{c}y}\right)^{2}
\nonumber
\\
&-|\beta|\left(-20k^{2}+Ee^{-A_{2}r_{c}y}+Fe^{-2A_{2}r_{c}y}\right)^{4}
\nonumber
\\
&=\left\lbrace -20k^{2}+\alpha \left(-20k^{2}\right)^{2}-|\beta|\left(-20k^{2}\right)^{4} \right\rbrace
+e^{-A_{2}r_{c}y}\left\lbrace E-40\alpha k^{2}E-4|\beta|\left(-20k^{2}\right)^{3}E \right\rbrace 
\nonumber
\\
&+e^{-2A_{2}r_{c}y}\left\lbrace F+\alpha E^{2}-40\alpha k^{2}F-6|\beta|E^{2}\left(-20k^{2}\right)^{2}-4|\beta|F\left(-20k^{2}\right)^{3} \right\rbrace
\nonumber
\\
&=P+Qe^{-A_{2}r_{c}y}+Re^{-2A_{2}r_{c}y}~.
\end{align}
%%%%%%%%%%%%%%%%%%%%%%%%%%%%%%%%%%%%%%%%%%%%%%%%%%%%%%%%%%%%%%%%%%%%%%%%%%
Thus the bulk gravitational action reads,
%%%%%%%%%%%%%%%%%%%%%%%%%%%%%%%%%%%%%%%%%%%%%%%%%%%%%%%%%%%%%%%%%%%%%%%%%%
\begin{align}
2\kappa _{5}^{2}\mathcal{A}&=\int d^{4}x\int dy r_{c}f(y)^{5/2}e^{-4A(y)}L
\nonumber
\\
&=\int d^{4}x\int dy r_{c} \left(1+A_{3}e^{-A_{2}r_{c}y}+A_{4}e^{-3A_{2}r_{c}y}\right)^{-5/3}e^{-4kr_{c}y}
\left(P+Qe^{-A_{2}r_{c}y}+Re^{-2A_{2}r_{c}y}\right)
\nonumber
\\
&=\int d^{4}x\int dy r_{c} e^{-4kr_{c}y}\left[P+\left\lbrace Q-\frac{5}{3}PA_{3} \right\rbrace e^{-A_{2}r_{c}y}
+\left\lbrace R-\frac{5}{3}QA_{3}+\frac{20}{9}PA_{3}^{2} \right\rbrace e^{-2A_{2}r_{c}y}\right]~.
\end{align}
%%%%%%%%%%%%%%%%%%%%%%%%%%%%%%%%%%%%%%%%%%%%%%%%%%%%%%%%%%%%%%%%%%%%%%%%%%
Integrating out the extra spatial coordinate, leads to a potential term for the radion field, which contributes only at $y=\pi$ and becomes,
%%%%%%%%%%%%%%%%%%%%%%%%%%%%%%%%%%%%%%%%%%%%%%%%%%%%%%%%%%%%%%%%%%%%%%%%%%
\begin{align}\label{App_Eq01}
2\kappa _{5}^{2}V(r_{c})&=e^{-4kr_{c}\pi}\left(\frac{P}{4k}\right)+e^{-4kr_{c}\pi}e^{-A_{2}r_{c}\pi}
\left\lbrace \frac{Q-\frac{5}{3}PA_{3}}{4k+A_{2}} \right\rbrace
\nonumber
\\
&+e^{-4kr_{c}\pi}e^{-2A_{2}r_{c}\pi}\left\lbrace \frac{R-\frac{5}{3}QA_{3}+\frac{20}{9}PA_{3}^{2}}{4k+2A_{2}}
\right\rbrace +\textrm{terms independent of}~r_{c}~.
\end{align}
%%%%%%%%%%%%%%%%%%%%%%%%%%%%%%%%%%%%%%%%%%%%%%%%%%%%%%%%%%%%%%%%%%%%%%%%%%
Such that the minima of the potential, leading to stabilized value of $r_{c}$ can be obtained by equating $\partial V/\partial r_{c}=0$ resulting in 
%%%%%%%%%%%%%%%%%%%%%%%%%%%%%%%%%%%%%%%%%%%%%%%%%%%%%%%%%%%%%%%%%%%%%%%%%%
\begin{align}
kr_{c}=\frac{16\alpha ^{2}}{9 \pi \sqrt{|\beta|}}\sqrt{-\frac{\Lambda \kappa _{5}^{2}}{6}}~\ln \left[\frac{2\left(R-\frac{5}{3}QA_{3}+\frac{20}{9}PA_{3}^{2}\right)}{\left(\frac{5}{3}PA_{3}-Q\right)+\sqrt{\left(Q-\frac{5}{3}PA_{3}\right)^{2}-4P\left(R-\frac{5}{3}QA_{3}+\frac{20}{9}PA_{3}^{2}\right)}} \right]~.
\end{align}
%%%%%%%%%%%%%%%%%%%%%%%%%%%%%%%%%%%%%%%%%%%%%%%%%%%%%%%%%%%%%%%%%%%%%%%%%%
The above expression can be enormously simplified by keeping terms to the leading order. In particular, one will have $P=-20k^{2}$, $Q=-(40/3)k^{2}A_{3}$ and $R=80kA_{1}A_{2}$ leading to \ref{Eq_04}, where we have assumed $A_{1}>k>A_{2}$ and $A_{1}>1$. It turns out that in order to have the stabilized value of $kr_{c}\simeq 12$, the above conditions are identically satisfied. Introduce the field $\phi$, such that the above potential becomes,
%%%%%%%%%%%%%%%%%%%%%%%%%%%%%%%%%%%%%%%%%%%%%%%%%%%%%%%%%%%%%%%%%%%%%%%%%%
\begin{align}
2\kappa _{5}^{2}V(\phi)&=\left(\frac{P}{4k}\right)\left(\frac{\phi}{\Phi}\right)^{4}+\left(\frac{\phi}{\Phi}\right)^{4+\frac{A_{2}}{k}}\left\lbrace \frac{Q-\frac{5}{3}PA_{3}}{4k+A_{2}} \right\rbrace
\nonumber
\\
&+\left(\frac{\phi}{\Phi}\right)^{4+2\frac{A_{2}}{k}}\left\lbrace \frac{R-\frac{5}{3}QA_{3}+\frac{20}{9}PA_{3}^{2}}{4k+2A_{2}}\right\rbrace +\textrm{terms independent of}~r_{c}~.
\end{align}
%%%%%%%%%%%%%%%%%%%%%%%%%%%%%%%%%%%%%%%%%%%%%%%%%%%%%%%%%%%%%%%%%%%%%%%%%%
This is the expression, which along with the previous identifications has been used to arrive at \ref{Eq_05}. Given the potential $V(\phi)$, one can compute,
%%%%%%%%%%%%%%%%%%%%%%%%%%%%%%%%%%%%%%%%%%%%%%%%%%%%%%%%%%%%%%%%%%%%%%%%%%
\begin{align}
2\kappa _{5}^{2}\frac{\partial V}{\partial \phi}&=\frac{1}{\Phi}\left(\frac{P}{4k}\right)\frac{4\phi ^{3}}{\Phi^{3}}
+\frac{1}{\Phi}\left(\frac{\phi}{\Phi}\right)^{3+\frac{A_{2}}{k}}\left(4+\frac{A_{2}}{k}\right)\left\lbrace \frac{Q-\frac{5}{3}PA_{3}}{4k+A_{2}} \right\rbrace
\nonumber
\\
&+\frac{1}{\Phi}\left(\frac{\phi}{\Phi}\right)^{3+2\frac{A_{2}}{k}}\left(4+2\frac{A_{2}}{k}\right)\left\lbrace \frac{R-\frac{5}{3}QA_{3}+\frac{20}{9}PA_{3}^{2}}{4k+2A_{2}}\right\rbrace ~,
\end{align}
%%%%%%%%%%%%%%%%%%%%%%%%%%%%%%%%%%%%%%%%%%%%%%%%%%%%%%%%%%%%%%%%%%%%%%%%%%
as well as,
%%%%%%%%%%%%%%%%%%%%%%%%%%%%%%%%%%%%%%%%%%%%%%%%%%%%%%%%%%%%%%%%%%%%%%%%%%
\begin{align}
2k\kappa _{5}^{2}\frac{\partial ^{2}V}{\partial \phi ^{2}}&=\frac{1}{\Phi^{2}}P\frac{3\phi ^{2}}{\Phi^{2}}
+\frac{1}{\Phi^{2}}\left(\frac{\phi}{\Phi}\right)^{2+\frac{A_{2}}{k}}\left(3+\frac{A_{2}}{k}\right)\left\lbrace Q-\frac{5}{3}PA_{3} \right\rbrace
\nonumber
\\
&+\frac{1}{\Phi^{2}}\left(\frac{\phi}{\Phi}\right)^{2+2\frac{A_{2}}{k}}\left(3+2\frac{A_{2}}{k}\right)\left\lbrace R-\frac{5}{3}QA_{3}+\frac{20}{9}PA_{3}^{2}\right\rbrace ~.
\end{align}
%%%%%%%%%%%%%%%%%%%%%%%%%%%%%%%%%%%%%%%%%%%%%%%%%%%%%%%%%%%%%%%%%%%%%%%%%%
Thus the minimum of the potential corresponds to $\partial V/\partial \phi=0$, 
%%%%%%%%%%%%%%%%%%%%%%%%%%%%%%%%%%%%%%%%%%%%%%%%%%%%%%%%%%%%%%%%%%%%%%%%%%
\begin{align}
P+\left(\frac{\phi}{\Phi}\right)^{\frac{A_{2}}{k}}\left\lbrace Q-\frac{5}{3}PA_{3} \right\rbrace+\left(\frac{\phi}{\Phi}\right)^{2\frac{A_{2}}{k}}\left\lbrace R-\frac{5}{3}QA_{3}+\frac{20}{9}PA_{3}^{2}\right\rbrace=0~.
\end{align}
%%%%%%%%%%%%%%%%%%%%%%%%%%%%%%%%%%%%%%%%%%%%%%%%%%%%%%%%%%%%%%%%%%%%%%%%%%%
Thus the radion mass becomes,
%%%%%%%%%%%%%%%%%%%%%%%%%%%%%%%%%%%%%%%%%%%%%%%%%%%%%%%%%%%%%%%%%%%%%%%%%%
\begin{align}
m_{\phi}^{2}&=\frac{\partial ^{2}V}{\partial \phi ^{2}}(r_{c});\qquad \frac{\partial V}{\partial \phi}(r_{c}) =0
\nonumber
\\
&=\frac{1}{2k\kappa _{5}^{2}}\frac{1}{\Phi^{2}}\frac{\phi ^{2}}{\Phi^{2}}\Bigg[-3\left(\frac{\phi}{\Phi}\right)^{\frac{A_{2}}{k}}\left\lbrace Q-\frac{5}{3}PA_{3} \right\rbrace -3\left(\frac{\phi}{\Phi}\right)^{2\frac{A_{2}}{k}}\left\lbrace R-\frac{5}{3}QA_{3}+\frac{20}{9}PA_{3}^{2}\right\rbrace
\nonumber
\\
&+\left(\frac{\phi}{\Phi}\right)^{\frac{A_{2}}{k}}\left(3+\frac{A_{2}}{k}\right)\left\lbrace Q-\frac{5}{3}PA_{3} \right\rbrace
+\left(\frac{\phi}{\Phi}\right)^{2\frac{A_{2}}{k}}\left(3+2\frac{A_{2}}{k}\right)\left\lbrace R-\frac{5}{3}QA_{3}+\frac{20}{9}PA_{3}^{2}\right\rbrace \Bigg]~.
\end{align}
%%%%%%%%%%%%%%%%%%%%%%%%%%%%%%%%%%%%%%%%%%%%%%%%%%%%%%%%%%%%%%%%%%%%%%%%%%
This helps to obtain the radion mass as in \ref{Eq_06}.
%%%%%%%%%%%%%%%%%%%%%%%%%%%%%%%%%%%%%%%%%%%%%%%%%%%%%%%%%%%%%%%%%%%%%%%%%%%%%%%%
%%%%%%%%%%%%%%%%%%%%%%%%%%%%%%%%%%%%%%%%%%%%%%%%%%%%%%%%%%%%%%%%%%%%%%%%%%%%%%%%
%%%%%%%%%%%%%%%%%%%%%%%%%%%%%%%%%%%%%%%%%%%%%%%%%%%%%%%%%%%%%%%%%%%%%%%%%%%%%%%%
\subsection{Stabilization in scalar-tensor representation}

Since any $f(R)$ theory of gravity has a dual description it would be interesting to understand the situation presented above in the dual scalar-tensor theory as well. This will provide a similar setup to the original Goldberger-Wise scheme but with back-reaction included. In this case the gravitational action is given by \ref{ST_01}, with the following solution,
%%%%%%%%%%%%%%%%%%%%%%%%%%%%%%%%%%%%%%%%%%%%%%%%%%%%%%%%%%%%%%%%%%%%%%%%%%
\begin{equation}
ds^{2}=e^{-2A(y)}\eta _{\mu \nu}dx^{\mu}dx^{\nu}+r_{c}^{2}dy^{2};\qquad A(y)=A_{0}+kr_{c}y+\frac{\kappa _{5}^{2}}{12}\psi _{0}^{2}e^{-A_{2}r_{c}y};\qquad \psi (y)=\psi _{0}e^{-\frac{A_{2}}{2}r_{c}y}
\end{equation}
%%%%%%%%%%%%%%%%%%%%%%%%%%%%%%%%%%%%%%%%%%%%%%%%%%%%%%%%%%%%%%%%%%%%%%%%%%
Note that the correspondence between Jordan and Einstein frame relates the parameter $A_{2}$ appearing here with that earlier. The Ricci scalar for the above metric can be evaluated leading to,
%%%%%%%%%%%%%%%%%%%%%%%%%%%%%%%%%%%%%%%%%%%%%%%%%%%%%%%%%%%%%%%%%%%%%%%%%%
\begin{align}
R&=\frac{1}{r_{c}^{2}}\left(8A''-20A'^{2}\right)
\nonumber
\\
&=-\frac{20}{9}a_{0}^{2}+\left(\frac{32b_{0}^{2}}{3\kappa _{5}^{2}}+\frac{40}{9}a_{0}b_{0} \right)\psi ^{2}-\frac{20}{9}b_{0}^{2}\psi ^{4}
\end{align}
%%%%%%%%%%%%%%%%%%%%%%%%%%%%%%%%%%%%%%%%%%%%%%%%%%%%%%%%%%%%%%%%%%%%%%%%%%
where, $a_{0}$ is a constant to be identified with $3k$, while, $b_{0}=\kappa _{5}^{2}A_{2}/4$. Further one can obtain,
%%%%%%%%%%%%%%%%%%%%%%%%%%%%%%%%%%%%%%%%%%%%%%%%%%%%%%%%%%%%%%%%%%%%%%%%%%
\begin{align}
g^{\mu \nu}\partial _{\mu}\psi \partial _{\nu} \psi =\frac{1}{r_{c}^{2}}\left(\frac{\partial \psi}{\partial y}\right)^{2}
=\frac{A_{2}^{2}}{4}\psi _{0}^{2}e^{-A_{2}r_{c}y}
\end{align}
%%%%%%%%%%%%%%%%%%%%%%%%%%%%%%%%%%%%%%%%%%%%%%%%%%%%%%%%%%%%%%%%%%%%%%%%%%
Such that the action becomes,
%%%%%%%%%%%%%%%%%%%%%%%%%%%%%%%%%%%%%%%%%%%%%%%%%%%%%%%%%%%%%%%%%%%%%%%%%%
\begin{align}
\mathcal{A}&=\int d^{4}x\int dy r_{c}e^{-4A}\Bigg[\frac{1}{2\kappa _{5}^{2}}
\left\lbrace -\frac{20}{9}a_{0}^{2}+\left(\frac{2\kappa _{5}^{2}A_{2}^{2}}{3}+\frac{10}{9}\kappa _{5}^{2}a_{0}A_{2}\right)\psi _{0}^{2}e^{-A_{2}r_{c}y}-\frac{5}{36}\kappa _{5}^{4}A_{2}^{2}\psi _{0}^{4}e^{-2A_{2}r_{c}y} \right\rbrace
\nonumber
\\
&-\frac{A_{2}^{2}}{8}\psi _{0}^{2}e^{-A_{2}r_{c}y}
-\left\lbrace \left(-\Lambda-\frac{2a_{0}^{2}}{3\kappa _{5}^{2}}\right)
+\left(\frac{A_{2}^{2}}{32}+\frac{a_{0}A_{2}}{3} \right)\psi ^{2}-\left(\frac{\kappa _{5}^{2}A_{2}^{2}}{24}\right)\psi ^{4} \right\rbrace -\Lambda\Bigg]
\end{align}
%%%%%%%%%%%%%%%%%%%%%%%%%%%%%%%%%%%%%%%%%%%%%%%%%%%%%%%%%%%%%%%%%%%%%%%%%%
Hence the potential becomes (except for terms independent of $r_{c}$) by introducing $R=\Phi e^{-kr_{c}\pi}$,
%%%%%%%%%%%%%%%%%%%%%%%%%%%%%%%%%%%%%%%%%%%%%%%%%%%%%%%%%%%%%%%%%%%%%%%%%%
\begin{align}
V(R)&=\frac{R^{4}}{\Phi^{4}}
\Bigg[\left(-\frac{4}{9}\frac{a_{0}^{2}}{\kappa _{5}^{2}}\right)\frac{1}{4k}
+\left(\frac{17}{96}A_{2}^{2}+\frac{2a_{0}A_{2}}{9}\right)\psi _{0}^{2}\frac{(R/\Phi)^{A_{2}/k}}{4k+A_{2}}
+\left(-\frac{\kappa _{5}^{2}A_{2}^{2}}{36}\right)\psi _{0}^{4}\frac{(R/\Phi)^{2A_{2}/k}}{4k+2A_{2}}\Bigg]
\end{align}
%%%%%%%%%%%%%%%%%%%%%%%%%%%%%%%%%%%%%%%%%%%%%%%%%%%%%%%%%%%%%%%%%%%%%%%%%%%
In this case as well the minima of the potential originates from the following algebraic equation,
%%%%%%%%%%%%%%%%%%%%%%%%%%%%%%%%%%%%%%%%%%%%%%%%%%%%%%%%%%%%%%%%%%%%%%%%%%%
\begin{align}
\left(-\frac{4}{9}\frac{a_{0}^{2}}{\kappa _{5}^{2}}\right)
+\left(\frac{17}{96}A_{2}^{2}+\frac{2a_{0}A_{2}}{9}\right)\psi _{0}^{2}(R/\Phi)^{A_{2}/k}
+\left(-\frac{\kappa _{5}^{2}A_{2}^{2}}{36}\right)\psi _{0}^{4}(R/\Phi)^{2A_{2}/k}=0
\end{align}
%%%%%%%%%%%%%%%%%%%%%%%%%%%%%%%%%%%%%%%%%%%%%%%%%%%%%%%%%%%%%%%%%%%%%%%%%%%
Leading to a solution for the stabilized potential, which only differs within the logarithm and hence have identical leading order behavior. The mass as well as the coupling with standard model fields work out in an identical fashion. Thus the two frames lead to very much similar expressions for the stabilized value of the radion field as well as its mass and couplings with standard model fields. Hence at least in this context the two frames have nearly identical physical behavior. 
%%%%%%%%%%%%%%%%%%%%%%%%%%%%%%%%%%%%%%%%%%%%%%%%%%%%%%%%%%%%%%%%%%%%%%%%%%%%%%%%%%%%%%%%%%%%%%%%%%%
%%%%%%%%%%%%%%%%%%%%%%%%%%%%%%%%%%%%%%%%%%%%%%%%%%%%%%%%%%%%%%%%%%%%%%%%%%%%%%%%%%%%%%%%%%%%%%%%%%%
%%%%%%%%%%%%%%%%%%%%%%%%%%%%%%%%%%%%%%%%%%%%%%%%%%%%%%%%%%%%%%%%%%%%%%%%%%%%%%%%%%%%%%%%%%%%%%%%%%%
%Bibliography
\bibliography{Gravity_1_full,Gravity_2_full,Gravity_3_partial,Brane,My_References}

\bibliographystyle{./utphys1}
%%%%%%%%%%%%%%%%%%%%%%%%%%%%%%%%%%%%%%%%%%%%%%%%%%%%%%%%%%%%%%%%%%%%%%%%%%%%%%%%%%%%%%%%%%%%%%%%%%%
%%%%%%%%%%%%%%%%%%%%%%%%%%%%%%%%%%%%%%%%%%%%%%%%%%%%%%%%%%%%%%%%%%%%%%%%%%%%%%%%%%%%%%%%%%%%%%%%%%%
%%%%%%%%%%%%%%%%%%%%%%%%%%%%%%%%%%%%%%%%%%%%%%%%%%%%%%%%%%%%%%%%%%%%%%%%%%%%%%%%%%%%%%%%%%%%%%%%%%%
\end{document}